\documentclass[]{spie}

\renewcommand{\baselinestretch}{1.00} 
 
\usepackage{standalone}
\usepackage{comment}

\usepackage{cite}
\usepackage{algorithmic}
\usepackage{amsmath,amssymb,amsfonts}
\usepackage{graphicx}
\graphicspath{ {./images/} {./figures/} }

\usepackage{subcaption}
\usepackage{pgf}
\usepackage{pgfplots}
\pgfplotsset{width=10cm,compat=1.14}
\usepgfplotslibrary{external}

\usepackage{textcomp}
\usepackage{xcolor}
\usepackage{datetime}

\usepackage{placeins}

\usepackage{tikz}
\usepackage{tikz-3dplot}
\usetikzlibrary{calc,arrows.meta,positioning,backgrounds}
\usepackage{setspace}

\title{Performance Estimation of a Real-Time Rosette Imager}

\author[a, b]{Gene Stoltz}
\author[c]{Marnus Stoltz}
\affil[a]{University of Johannesburg, South Africa}
\affil[b]{Council for Scientific and Industrial Research of South Africa}
\affil[c]{University of Otago, New Zealand}

\authorinfo{Further author information: (Send correspondence to G.Stoltz)\\G.Stoltz: E-mail: ggsgene@gmail.com}

\begin{document}
\maketitle

\begin{abstract}
In this paper, we model a real-time feasible rosette imager, consisting of a rosette scanner, an optical sensor and a deterministic image reconstruction algorithm. We fine-tune the rosette imager through selecting the appropriate sensor field of view and rosette pattern. The sensor field of view is determined through a greedy approach using uniform random sampling. Furthermore, the optimal rosette pattern is selected by determining which pattern best covers the imaging area uniformly. 

We explore image sparsity, image decimation and Gaussian filtering in a well-known natural data set and dead leaves data set using the PSNR, Peak-Signal-to-Noise Ratio. This exploration helps to establish a connection between PSNR and image sparsity. Furthermore, we compare various rosette imager configurations in a Bayesian framework. We also conclude that the rosette imager does not outperform a focal-plane array of equivalent samples in terms of image quality but can match the performance.
\end{abstract}


\section{Introduction}
Infrared imaging is a key sensing component in support of digital technologies. Development of infrared imaging have shifted focus from scanning single detectors to focal-plane arrays, driven mostly by advances in manufacturing technology~\cite{IRTechBook}. Single detector scanning systems lack the quality and resolution of current focal-plane array systems and require advanced sampling strategies and higher data processing. The improvement of computational capability and development of new sampling theories, such as compressed sensing~\cite{donoho2006compressed}, has provided new opportunities for imaging through single detector scanning systems such as the rosette scanner.

It was shown that an infrared rosette scanner with a single detector, equivalent to an infrared seeker system, can produce images offline via compressed sensing~\cite{Uzeler2013}. The image formation process through compressed sensing was further improved by modelling detector measurements as the sum of multiple pixels, replacing the binary orthonormal measurement matrix first used~\cite{Uzeler2015}\cite{Uzeler2016}. Recent attempts to improve image reconstruction quality either combines multiple frames~\cite{Jiang2017} or smaller image blocks~\cite{Tong2018}. Here we design a real-time imaging device using a single detector scanning system that match the performance of focal-plane arrays.

The paper is structured as follow: In Section 2 we introduce the rosette imager and a performance optimisation process. Then, in Section 3, we discuss the methods used and experimental protocols followed for evaluating the optimised imager. The discussion includes a description of the testing data sets, Peak-Signal-to-Noise Ratio(PSNR) as performance measure and the importance of image sparsity for parameter inference. Thereafter, in Section 4, we discuss performance results of the rosette imager based on four rosette patterns and a focal-plane array. Lastly, in Section 5, we discuss future work and give some concluding remarks.


\section{The rosette imager}
The rosette imager is defined by the rosette scanning pattern and a compressed sensing setup.
Briefly, the rosette scanning pattern is the sampling pattern produced by the mechanical movement of an actual imager. The compressed sensing setup is how the samples are reconstructed to represent the image that was sampled.
We discuss these two components in more detail below. 
The requirements for the rosette imager being designed is as follow: 
\begin{itemize}
    \item An output image with resolution of $256\times256$ pixels.
    \item A frame rate of 25 frames per second~(fps).
    \item Real-time performance, subjected to a latency of 2 frames or $80~ms$. 
\end{itemize}
The designed imager must be practical and therefore the physical realisation can be achieved through utilising a rosette scanner. The rosette scanner has multiple parameters that can be adjusted to achieve different optimisations.  For example, sensor sensitivity in relation to integration times can be improved by using slower rotation frequencies and lower sampling rates~\cite{Tajime1980}. 
 
In this paper only the rosette scanning pattern and the detector field of view will be optimised for best image quality. It must be noted that all physical effects, such as lens distortion, mechanical precision, sensor noise and integration times, are ignored in the system evaluation.

\subsection{The Rosette Scanning Pattern}
\begin{figure}[h!t]
    \centering
    \begin{subfigure}{.35\columnwidth}
        \centering
        \fontsize{8pt}{10pt}\selectfont
        \def\svgwidth{0.9\columnwidth}
        \input{./figures/rosetteScannerTex.pdf_tex}
        \caption{Physical Layout}
        \label{fig:rosettescanner_layout}
    \end{subfigure}%
    \begin{subfigure}{.45\columnwidth}
        \vspace{0.8cm}
        \begin{minipage}{\linewidth}
        \centering
        \fontsize{8pt}{10pt}\selectfont
        \def\svgwidth{0.9\columnwidth}
\begingroup%
  \makeatletter%
  \providecommand\color[2][]{%
    \errmessage{(Inkscape) Color is used for the text in Inkscape, but the package 'color.sty' is not loaded}%
    \renewcommand\color[2][]{}%
  }%
  \providecommand\transparent[1]{%
    \errmessage{(Inkscape) Transparency is used (non-zero) for the text in Inkscape, but the package 'transparent.sty' is not loaded}%
    \renewcommand\transparent[1]{}%
  }%
  \providecommand\rotatebox[2]{#2}%
  \newcommand*\fsize{\dimexpr\f@size pt\relax}%
  \newcommand*\lineheight[1]{\fontsize{\fsize}{#1\fsize}\selectfont}%
  \ifx\svgwidth\undefined%
    \setlength{\unitlength}{167.39099337bp}%
    \ifx\svgscale\undefined%
      \relax%
    \else%
      \setlength{\unitlength}{\unitlength * \real{\svgscale}}%
    \fi%
  \else%
    \setlength{\unitlength}{\svgwidth}%
  \fi%
  \global\let\svgwidth\undefined%
  \global\let\svgscale\undefined%
  \makeatother%
  \begin{picture}(1,0.81337449)%
    \lineheight{1}%
    \setlength\tabcolsep{0pt}%
    \put(0,0){\includegraphics[width=\unitlength,page=1]{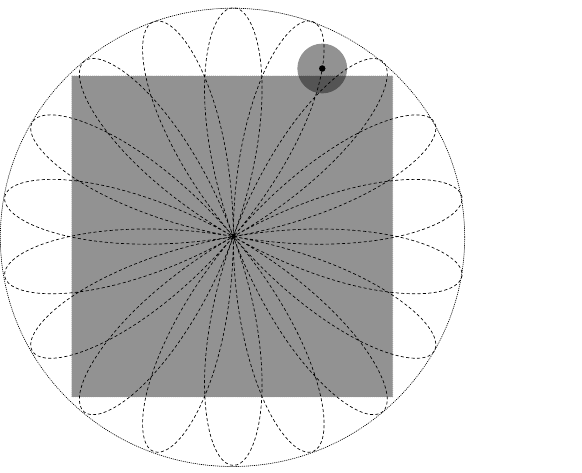}}%
    \put(0.84416232,0.07834158){\color[rgb]{0,0,0}\makebox(0,0)[lt]{\lineheight{1.25}\smash{\begin{tabular}[t]{l}FOV\end{tabular}}}}%
    \put(0.82620711,0.71018504){\color[rgb]{0,0,0}\makebox(0,0)[lt]{\lineheight{1.25}\smash{\begin{tabular}[t]{l}IFOV\end{tabular}}}}%
    \put(0.64116105,0.7770635){\color[rgb]{0,0,0}\makebox(0,0)[lt]{\lineheight{1.25}\smash{\begin{tabular}[t]{l}Sampling Point\end{tabular}}}}%
    \put(0.64419747,0.00987113){\color[rgb]{0,0,0}\makebox(0,0)[lt]{\lineheight{1.25}\smash{\begin{tabular}[t]{l}Imaging Area\end{tabular}}}}%
    \put(0,0){\includegraphics[width=\unitlength,page=2]{rosettePattern.pdf}}%
  \end{picture}%
\endgroup%

        \end{minipage}
        \vspace{0.6cm}
        \caption{Scanning Pattern Example}
        \label{fig:rosettescanner_pattern}
    \end{subfigure}    
    \caption{The Rosette Scanner. A wedge typed scanner is demonstrated in Figure~\ref{fig:rosettescanner_layout}. A typical scanning pattern with frequency $f_1=100Hz$ and $f_2=-175Hz$ (n,m = 11,4) is shown in Figure~\ref{fig:rosettescanner_pattern}. The system's Field of View~(FOV) is the total area covered by the rosette scanner. The Instantaneous Field of View~(IFOV) is the field the sensor covers while the sampling point indicates the optical axes of the detector. }
    \label{fig:rosettescanner}
\end{figure}

The rosette scanning pattern can be realised by using a combination of rotating canted mirrors, optic wedges or canted rotating detector. The paper will use a wedge typed rosette scanner as shown in Figure~\ref{fig:rosettescanner}. The rosette scanner must satisfy certain condition to successfully form a scanning pattern. The scanning pattern is presented by the locus that the Instantaneous Field of View~(IFOV) will travel when projected onto a plane perpendicular to the rotation axis (See Appendix~\ref{app:rosette} for governing equations). The different rosette scanning patterns are denoted by two integer values $(n, m)$. 

\FloatBarrier

The physical realisation of a rosette scanner have various processes that impose limitations such as manufacturing quality and cost. The following limitations can be imposed on the rosette scanner parameters used for the design of the rosette imager.
\begin{itemize}
\item The rotational frequency of the mirror and wedge is limited to 2500~Hz.
\item The maximum sampling frequency of the detector is limited to 510~kHz.
\end{itemize}

\subsection{The Compressed Sensing Setup}
The compressed sensing setup consist of a measurement and reconstruction matrix ~\cite{foucart2013mathematical}. The measurement matrix is governed by the limitations of the rosette scanner while the reconstruction matrix, generated by the reconstruction algorithm, governs the speed of image formation and image quality.

\begin{figure}[h!t]
    \centering
    \begin{subfigure}[b]{.39\columnwidth}    
        \centering
    \includegraphics[width=0.5\linewidth,interpolate=false]{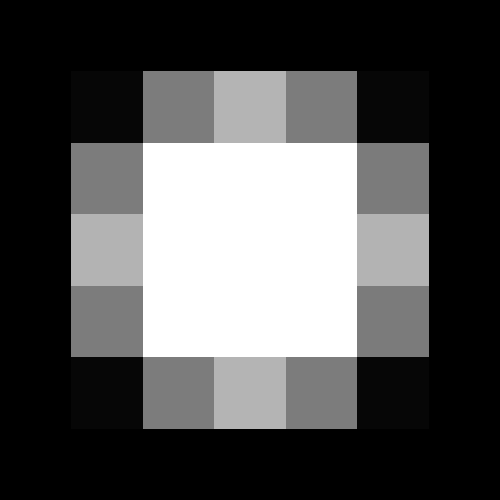}
    \end{subfigure}
    \caption{A probe with a radius of 2 pixels, starting in the middle of pixel (4, 4). The white pixels indicate~a~1 and the black pixels indicate~a~0 with the fractional values of the circle indicated by the grey values.}
    \label{fig:probe_example}
\end{figure}

The measurement matrix represents the detector output as a sum of pixel intensity values related to the output image~\cite{Uzeler2016}. The sample positions within the rosette pattern is discretised such that the optical axis of the detector is always in the middle of a pixel within the output image. The IFOV is dicretized in-terms of the output image and can be measured in pixel size. An example of a discretized detector is shown in Figure~\ref{fig:probe_example}. The discretized IFOV is placed on the sample position of the rosette pattern. The generated image is flattened and coincide with a row entry in the measurement matrix. For every sample in the rosette scanning pattern a row is added into the measurement matrix. It must be noted that the image formation process assumes a discrete world where the mapping from a continuous to a discrete domain has not been taken into account. The reference to the IFOV might be misleading when talking about the imaging domain, thus it will rather be referred to as the probe size.

The reconstruction matrix is based on the chosen image reconstruction algorithm.
The choice of reconstruction algorithm should allow for real-time execution while maintaining appropriate quality. A fair amount of reconstruction algorithms exist such as Basis Pursuit Denoising (BPDN)~\cite{BergFriedlander2008} and total variation regularisation (NESTA). Unfortunately, these algorithms are not suitable for real-time applications. A good candidate is the Fourier Domain Regularisation Inversion (FDRI) algorithm developed in 2018 by Czajkowski et al and has equivalent performance to NESTA~\cite{Czajkowski2018}. The FDRI algorithm is based on the calculation of the Moore-Penrose pseudoinverse. The FDRI is not an iterative algorithm and allow reconstruction by performing a single matrix multiplication with a predetermined reconstruction matrix. The reconstruction matrix can be calculated before the execution of imaging, but require the knowledge of a known measurement matrix. In case of a rosette scanning pattern, the measurement matrix is known beforehand. It is important to note that FDRI is not an $l1$-regularisation algorithm but approximate the effects.

The real-time expectation of using the FDRI reconstruction matrix can be calculated by assuming an output image of $256\times256$ with $20\%$ samples. The reconstruction matrix of size $65536\times 13107$ can be multiplied by the measurement vector of size $13107$ within $20ms$ utilising a GPU. (Tested on a nVidia GeForce 1060). The estimated speed of the execution makes the reconstruction algorithm real-time in-terms of the expected frame rate.

\subsection{Design Optimisations}
The rosette imager can now be designed through the optimisation of the probe size and the rosette scanning pattern. The probe size and sample positions have direct influence on the constructed measurement matrix and influencing the coherence between the measurement and reconstruction matrices. Although, the application of compressed sensing theory is limited by the fact that the probe samples in the spatial domain which is equivalent to the output domain. Despite this, compressed sensing principles will still be applied and random uniform sampling would be assumed to provide optimal reconstruction possibility.

\begin{figure}[ht]
    \centering
    \begin{subfigure}[b]{.4\columnwidth}
        \centering
        \includegraphics[width=0.99\linewidth]{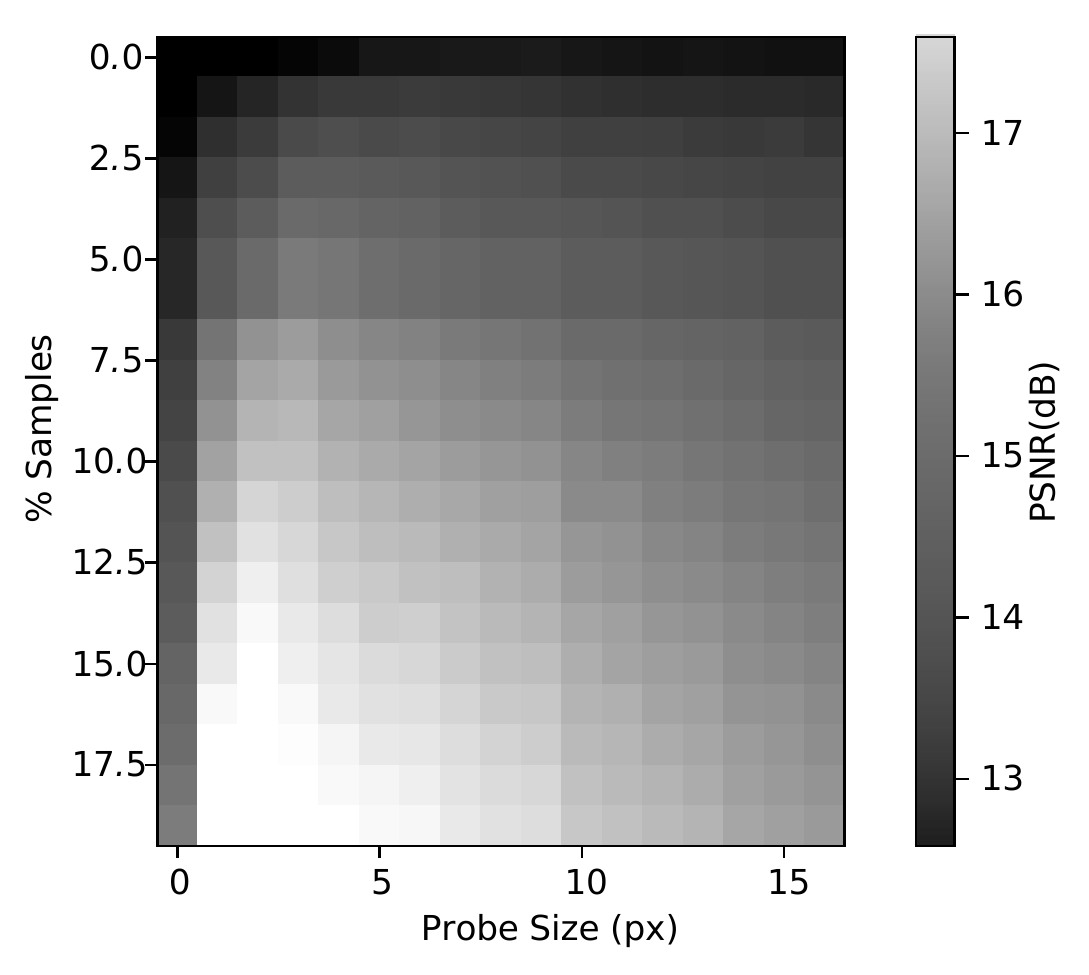}
        \caption{The percentage samples versus probe size with grey values indicating the relative PSNR values.}
        \label{fig:probe_optimise}
    \end{subfigure}%
    \begin{subfigure}[b]{.4\columnwidth}
        \centering
        \includegraphics[width=0.99\linewidth]{./figures/graph_rtr_probe.tex}
        \caption{PSNR versus probe size for 19.7\% samples. A probe size of 2.25 pixels provided the maximum PSNR value.}
        \label{fig:probe_sampling_vesus_ifov}
    \end{subfigure}%
    \caption{The result summary for the probe size optimisation process.}
    \label{fig:sampling_vesus_ifov_samples}
\end{figure}

To find the optimal probe size, a greedy approach is followed. The probe size and the percentage samples are linearly scanned while measuring the average PSNR over the dead leaves data set. For every probe size and sample percentage a set of random coordinates are uniformly distributed across the image. The number of generated coordinates in the set coincide with the sample percentage. The results of the greedy approach is shown in Figure~\ref{fig:probe_optimise}. 

Interpreting the results it can be concluded that a small probe, provide a higher maximum PSNR but require a higher percentage of samples. For a certain sampling percentage there exist an optimal probe size which will produce the maximum possible PSNR performance while applying uniform random sampling. Combining the compressed sensing theory and evidence from these experimental results the optimal performance can be achieved by minimising the number of pixels not part of a sampled probe and minimising the amount of probe overlap.

The rosette imager design will have a single sample percentage and probe size. The sample percentage can be estimated by assuming the samples within the field of view of the rosette scanning pattern are uniformly distributed. A square within a circle will consist of approximately $63.6\%$ of the circle area. A detector sampling at 510kHz will produce $20400$~samples per $40$~ms, thus $20400$ samples per rosette scanning pattern at 25fps. Approximately $12974$ samples will fall within the imaging area and with an output image $256\times256$ a sample percentage of $19.7\%$ will be achieved. Figure~\ref{fig:probe_sampling_vesus_ifov} provides the result related to 19.7\% sampling and produces probe size of 2.25 pixels at the maximum PSNR.

\begin{figure}[ht]
    \centering
    \begin{subfigure}[b]{.4\columnwidth}
        \centering
        \includegraphics[width=0.9\linewidth]{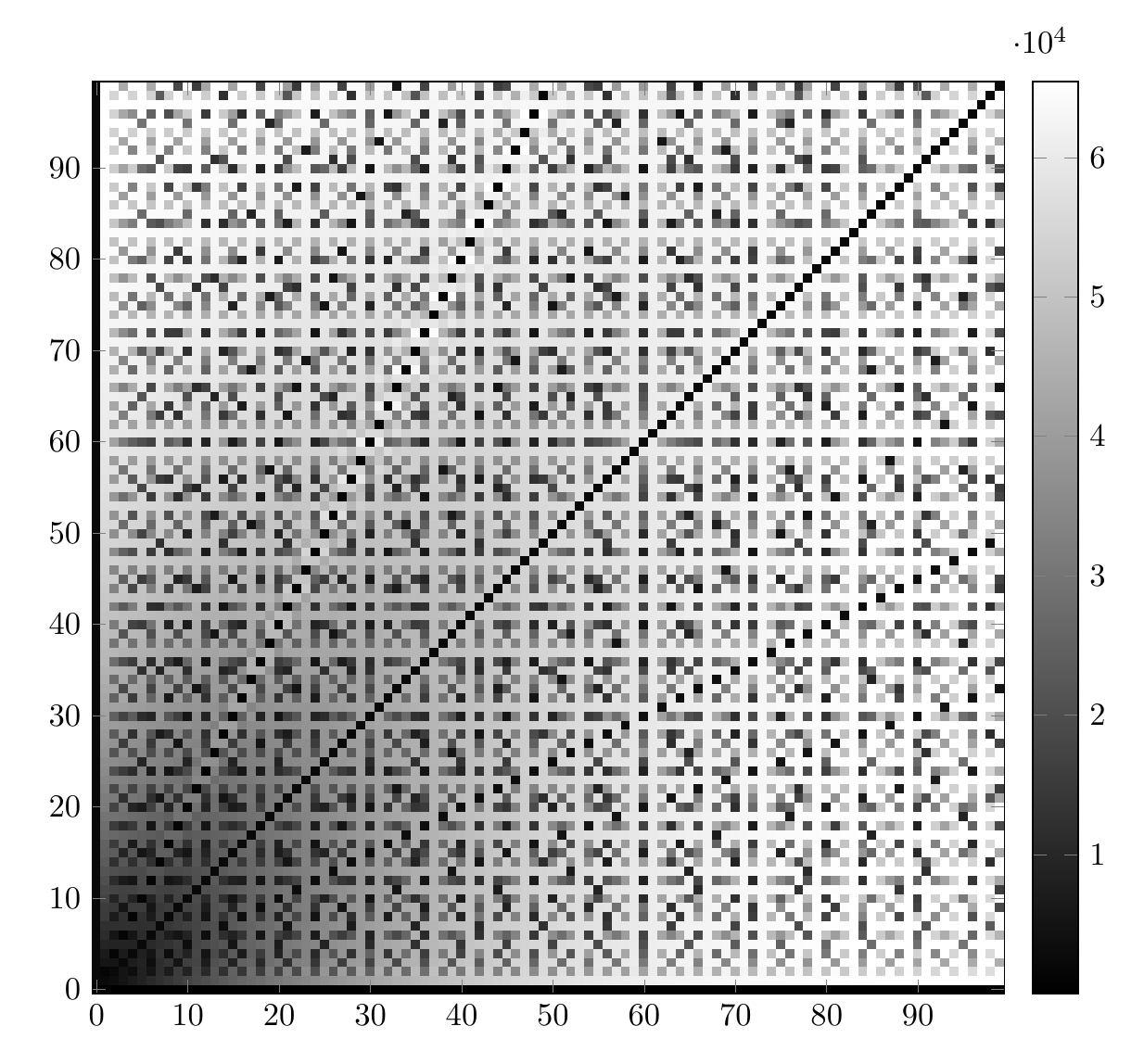}
        \caption{$(n,m)$ with plotted amount of non-zeros.}
        \label{fig:optimise1}
    \end{subfigure}%
    \begin{subfigure}[b]{.4\columnwidth}
        \centering
        \includegraphics[width=0.9\linewidth]{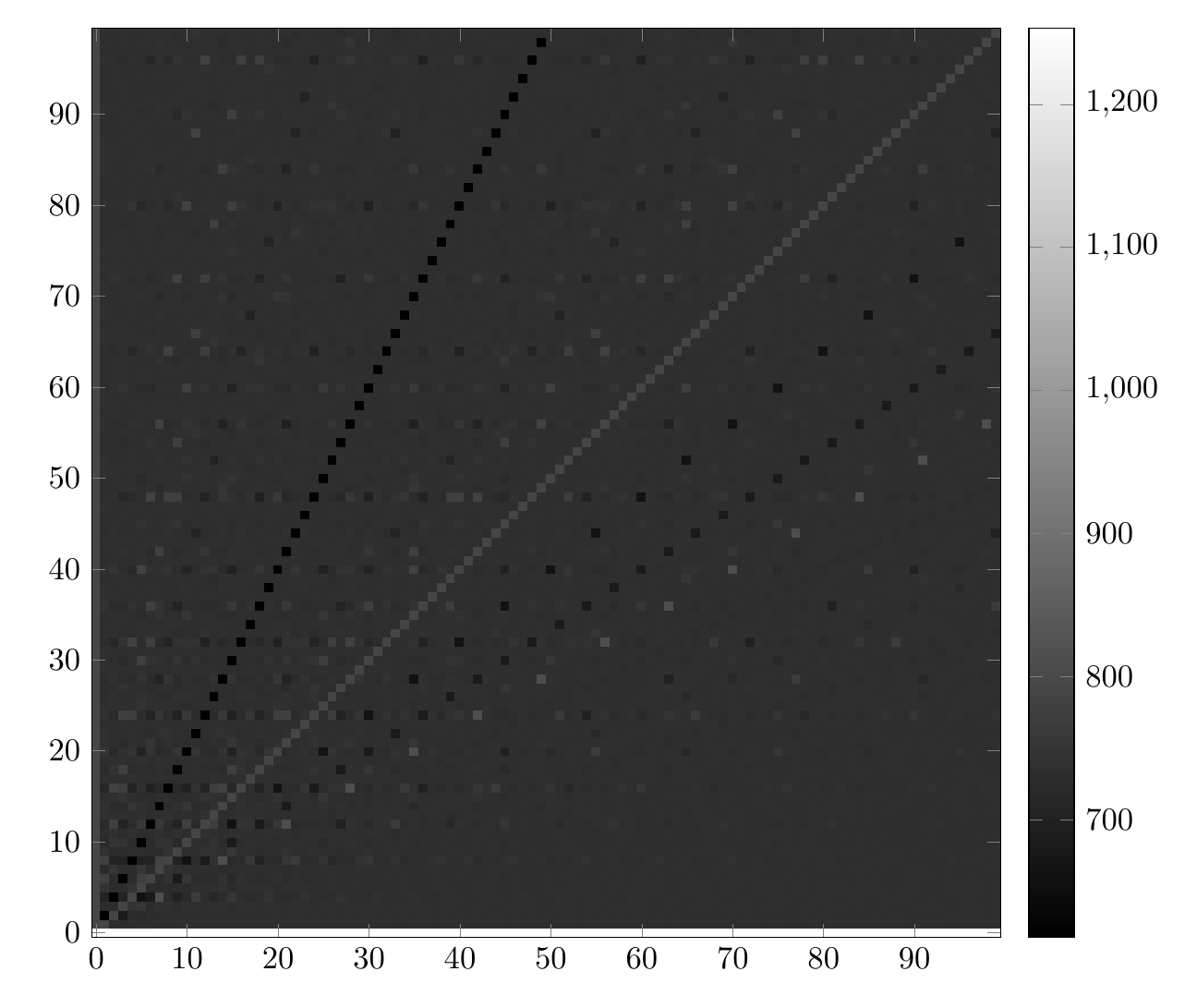}
        \caption{$(n,m)$ with plotted average bin size of the pixels.}
        \label{fig:optimise_max1}
    \end{subfigure}
    \begin{subfigure}[b]{.4\columnwidth}
        \centering
        \includegraphics[width=0.9\linewidth]{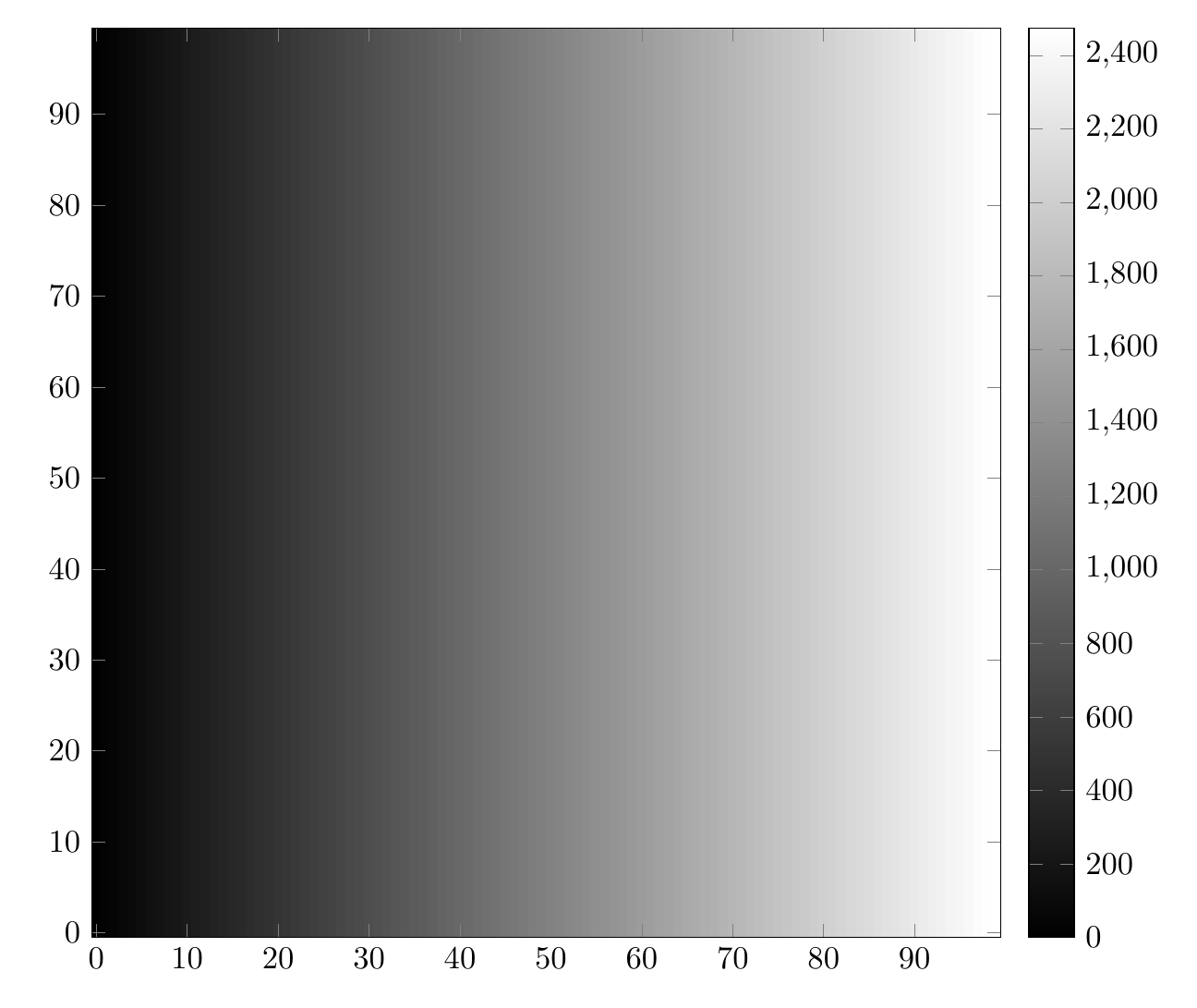}
        \caption{$(n,m)$ with plotted frequencies for $f_2$.}
        \label{fig:optimise}
    \end{subfigure}%
    \begin{subfigure}[b]{.4\columnwidth}
        \centering
        \includegraphics[width=0.9\linewidth]{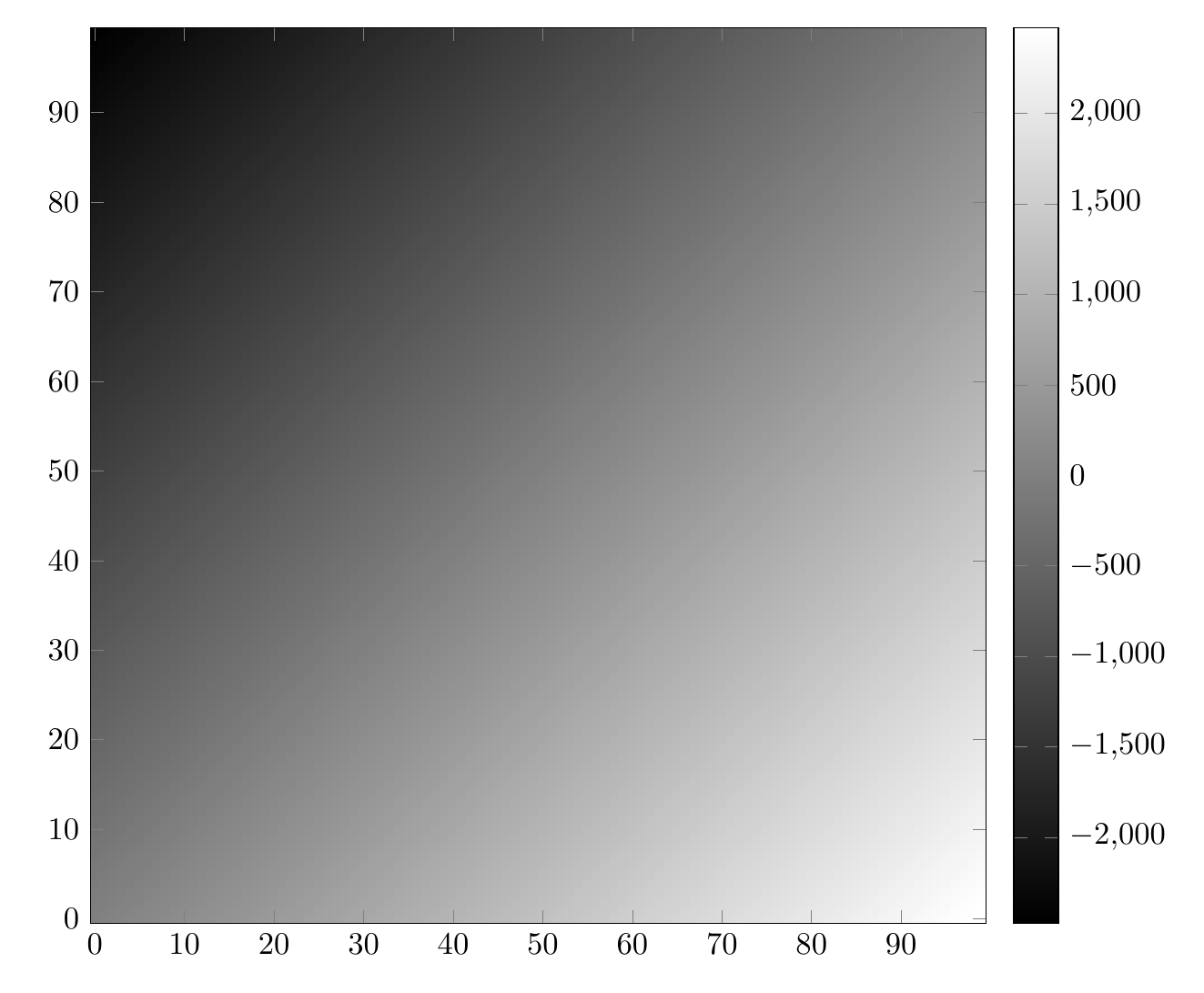}
        \caption{$(n,m)$ with plotted frequencies for $f_2$.}
        \label{fig:optimise_max}
    \end{subfigure}    
    \caption{Optimisation Images}
\end{figure}

In order to find the optimal scanning pattern we set a criteria based on minimal probes overlap and maximal number of pixels scanned within the imaging area. The probe overlap was already minimised by selecting the optimal probe size. To determine the optimal scanning pattern a 2-dimensional histogram is constructed using the pixels as bins and filling the bins with the probe positioned by the selected rosette pattern. The number of non-zero bins are counted while the pattern producing the most non-zero pixels are assumed to cover the imaging area uniformly. The rosette scanner parameters $(n,m)$ are linearly searched from $0-1000$ with the amount of non-zero entries plotted in Figure~\ref{fig:optimise}. The maximum value of the non-zero entries was measured and all entries equal to the maximum are plotted in Figure~\ref{fig:optimise_max} (m and n only range from 0-100 for display purposes). 

From the million n,m-values ($1000\times1000$) $1150$ $(m, n)$ combinations provided equivalent global maximum values of non-zero elements. The values were further filtered by $(n,m)$ combinations providing rotational frequencies below 2500~Hz and 1000~Hz. The top three patterns for each filter are shown in Table~\ref{tab:optimising_sets} and Table~\ref{tab:optimising_sets2}.

\begin{table}[ht]
    \centering
    \caption{Top 3 rosette pattern producing most non-zero value with rotation frequencies less than 2500~Hz}
    \label{tab:optimising_sets}
\begin{tabular}{|c|c|c|c|c|c|}
non-zeros & n & m & Frequency 1 & Frequency 2 & Number of Samples\\
\hline
65450 & 8 &99 &2475 & 2275 & 11966\\
65428 & 15&98 &2450 & 2075  & 11977\\
65444 & 20&97 & 2425& 1925 & 11976\\
\end{tabular}
\end{table}

\begin{table}[ht]
    \centering
    \caption{Top 3 rosette pattern producing most non-zero value with rotation frequencies less than 1000~Hz}
    \label{tab:optimising_sets2}
\begin{tabular}{|c|c|c|c|c|c|}
non-zeros & n & m & Frequency 1 & Frequency 2 & Number of Samples\\
\hline
64383 & 76 & 37 &925 & -975  & 12008\\
64377 & 76 & 39 &975 & -925  & 12009\\
63877 & 72 & 35 & 825& -925  & 11946\\
\end{tabular}
\end{table}

\begin{figure}[ht]
    \centering
    \begin{subfigure}[b]{.33\columnwidth}
        \centering
        \includegraphics[width=0.9\linewidth]{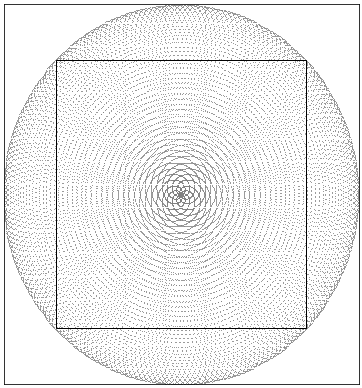}
        \caption{(n,m) = (8,99)}
    \end{subfigure}%
    \begin{subfigure}[b]{.33\columnwidth}
        \centering
        \includegraphics[width=0.9\linewidth]{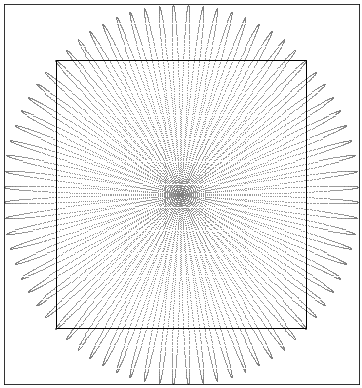}
        \caption{(n,m) = (76,37)}
    \end{subfigure}%
    \begin{subfigure}[b]{.33\columnwidth}
        \centering
        \includegraphics[width=0.9\linewidth]{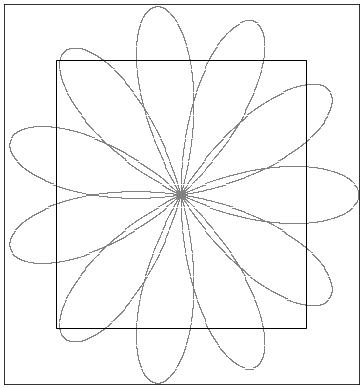}
        \caption{(n,m) = (11,4)}
        \label{fig:pattern_11_4}
    \end{subfigure}
    \begin{subfigure}[b]{.33\columnwidth}
        \centering
        \includegraphics[width=0.9\linewidth]{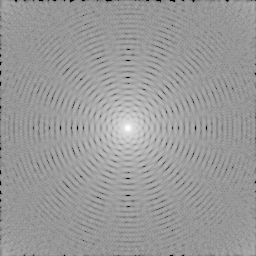}
        \caption{(n,m) = (8,99),d=2.25 }
    \end{subfigure}%
    \begin{subfigure}[b]{.33\columnwidth}
        \centering
        \includegraphics[width=0.9\linewidth]{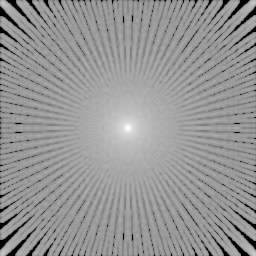}
        \caption{(n,m) = (76,37),d=2.25}
    \end{subfigure}%
    \begin{subfigure}[b]{.33\columnwidth}
        \centering
        \includegraphics[width=0.9\linewidth]{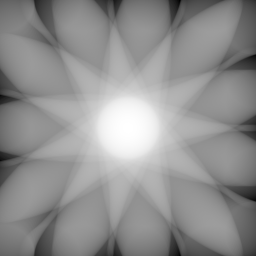}
        \caption{(n,m) = (11,4),d=31}
    \end{subfigure}        
    \caption{Illustration of the optimal sampling patterns with limiting frequencies}
    \label{fig:sampling_patterns}
\end{figure}

Figure~\ref{fig:sampling_patterns} provide an illustration of the best 2500~Hz and 1000~Hz pattern with their respective histograms with probe size equal to 2.25 pixels. The pattern shown in Figure~\ref{fig:pattern_11_4} is the pattern used for most previous publications with $f1=100Hz$ and $f2=-175Hz$. The probe size was adjusted such that the probe cover all pixels, equivalent to selecting the optimal IFOV for an infrared seeker system. The radius was found to be 31 pixels. The recommended sampling patterns for 2500~Hz are (8, 99) and for 1000~Hz are (76, 37) with a probe size of 2.25 pixels respectively.

\section{Assessment Methods and Protocol}
\label{sec:evaluation}
\subsection{Data Sets}
\label{sec:datasets}
\begin{figure}[h!t]
    \centering
    \begin{subfigure}{.33\columnwidth}
        \centering
        \includegraphics[width=0.9\linewidth]{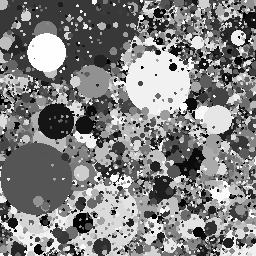}
    \end{subfigure}%
    \begin{subfigure}{.33\columnwidth}
        \centering
        \includegraphics[width=0.9\linewidth]{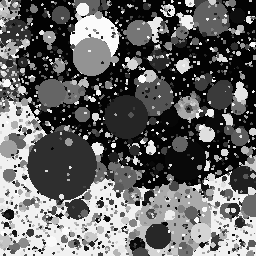}
    \end{subfigure}%
    \begin{subfigure}{.33\columnwidth}
        \centering
        \includegraphics[width=0.9\linewidth]{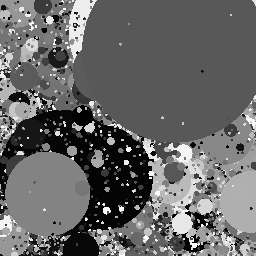}
    \end{subfigure}%
    \caption{Dead leaves data set image examples.}
    \label{fig:deadleaves}
\end{figure}

In the assessment of the rosette imager we used two datasets namely, the dead leaves dataset and the natural scene dataset. The deadleaves dataset is significant due to its scale invariance. Lee~et~el~\cite{Lee2001} introduced a scale invariance case of the general dead leaves model. This scale invariant case occurs when the power law distribution parameter of the simulation is set to 3. The scale invariant case of the dead leaves model approximate natural scenes when a small size cut-off on the circle radius's are introduced~\cite{Gousseau2008}. Perfect simulation of the stochastic geometry occurs~\cite{Kendall1999} when backwards construction of the pattern is performed successfully simulating such an image. Generated dead leaves image samples are shown in Figure~\ref{fig:deadleaves}.

\begin{figure}[h!t]
    \centering
    \begin{subfigure}{.33\columnwidth}
        \centering
        \includegraphics[width=0.9\linewidth]{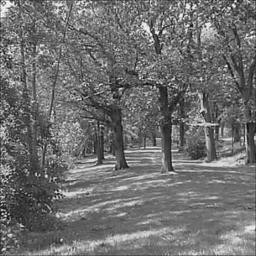}
        \caption{Forest}
    \end{subfigure}%
    \begin{subfigure}{.33\columnwidth}
        \centering
        \includegraphics[width=0.9\linewidth]{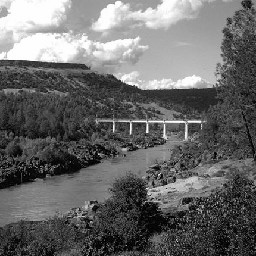}
        \caption{Open Country}
    \end{subfigure}%
    \begin{subfigure}{.33\columnwidth}
        \centering
        \includegraphics[width=0.9\linewidth]{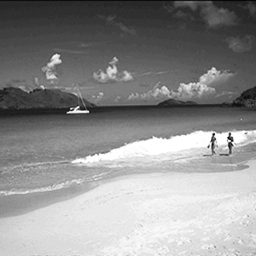}
        \caption{Coast}
    \end{subfigure}%
    \caption{Natural scene imaging examples}
    \label{fig:natural}
\end{figure}

The natural scene data set is divided into various categories \cite{oliva2001modeling} such as forest, coast, mountains and open country. Examples are given in Figure~\ref{fig:natural}.

\subsection{Peak-Signal-to-Noise Ratio}
Reconstructed images from the data sets (described above) are assessed based on the Peak-Signal-to-Noise Ratio (PSNR) measure. The PSNR measure is the inverse logarithm of the Mean Square Error between two discrete valued images (See Appendix~\ref{app:psnr}). Images are converted to 16~bit, before measuring the PSNR, placing the maximum PSNR value at $96.33$~dB. It is common to use the Structural Similarity(SSIM)~\cite{Z.WangA.C.Bovik2004} measure when comparing images although it is interesting to note that SSIM is related to PSNR~\cite{Hore2010, Dosselmann2011, Brunet2012}. It can be argued that according to some research papers SSIM is a better metric compared to PSNR or MSE. Not SSIM or PSNR can be used as a norm and is thus not an adequate measurement for vector spaces although MSE can be calculated from PSNR. SSIM might indicate the likelihood that the human visual system would find certain images more visually pleasing but this is not the purpose of a sampling system. The purpose is to accurately reconstruct a sampled signal and measure the accuracy. Therefore the PSNR measure is adopted and is also proposed for future research.

\subsection{Image Sparsity}
\label{sec:sparsity}
In compressed sensing one of the factors determining the quality of signal reconstruction is the sparsity of the signal itself. A more sparse signal will require fewer samples to reconstruct at a higher accuracy. The expected PSNR values should reflect this effect and is explored in Section~\ref{sec:dct_sparse} using the Discrete Cosine Transform~(DCT). Also in the section, the large histogram variances on the PSNR values assert the requirement for advanced statistical techniques and the dead leaves data set is shown to be adequate evaluation set for comparing imaging algorithms. 

Following on image sparsity in the DCT, an image with a fair amount of detail should reconstruct worse than an image with less detail. The dead leaves data set is generated to represent a highly detailed natural image, while the natural scene database consist of landscapes (low detail), mountain areas (medium detail) and forests(high detail). In Section~\ref{sec:fpa_sparsity} image sparsity is related to image detail.

\subsubsection{Image sparsity in the Discrete Cosine Transform}
\label{sec:dct_sparse}
The sparsity of an image is measured in accordance to a specific transformation basis. The Discrete Cosine Transform~(DCT) will be used as the sparsity basis. The image sparsity is taken as the percentage of non-zero coefficients within the transform. Sparse images was created by selecting a percentage of the highest coefficients in the transform, setting all other coefficients to zero and taking the inverse transform. The number of coefficients required to reproduce an image at maximum PSNR value indicates the sparsity of the image. 

\begin{figure}[h!t]
    \centering
    \begin{subfigure}{.24\columnwidth}
        \centering
        \includegraphics[width=0.9\linewidth]{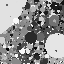}
        \caption*{Original\\ 100\%}
    \end{subfigure}%
    \begin{subfigure}{.24\columnwidth}
        \centering
        \includegraphics[width=0.9\linewidth]{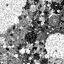}
        \caption*{PSNR: 20.40\\ 20\%}
    \end{subfigure}%
    \begin{subfigure}{.24\columnwidth}
        \centering
        \includegraphics[width=0.9\linewidth]{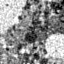}
        \caption*{PSNR: 17.95\\ 10\%}
    \end{subfigure}%
    \begin{subfigure}{.24\columnwidth}
        \centering
        \includegraphics[width=0.9\linewidth]{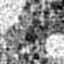}
        \caption*{PSNR: 16.40\\ 5\%}
    \end{subfigure}%
    \caption{An example image from the dead leaves database reconstructed at different sparsity percentages based on the Discrete Cosine Transform}
    \label{fig:sparse_natural}
\end{figure}

\begin{figure}[h!t]
    \centering
    \begin{subfigure}{.24\columnwidth}
        \centering
        \includegraphics[width=0.9\linewidth]{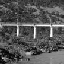}
        \caption*{Original\\ 100\%}
    \end{subfigure}%
    \begin{subfigure}{.24\columnwidth}
        \centering
        \includegraphics[width=0.9\linewidth]{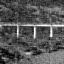}
        \caption*{PSNR: 25.89\\ 20\%}
    \end{subfigure}%
    \begin{subfigure}{.24\columnwidth}
        \centering
        \includegraphics[width=0.9\linewidth]{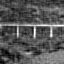}
        \caption*{PSNR: 23.13\\ 10\%}
    \end{subfigure}%
    \begin{subfigure}{.24\columnwidth}
        \centering
        \includegraphics[width=0.9\linewidth]{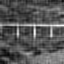}
        \caption*{PSNR: 21.21\\ 5\%}
    \end{subfigure}%
    \caption{Example image from the Scene database reconstructed at different sparsity percentages based on the Discrete Cosine Transform. The images was zoomed ($128\times128$) to allow for closer investigation.}
    \label{fig:sparse_deadleaves}
\end{figure}

\begin{figure}[h!t]
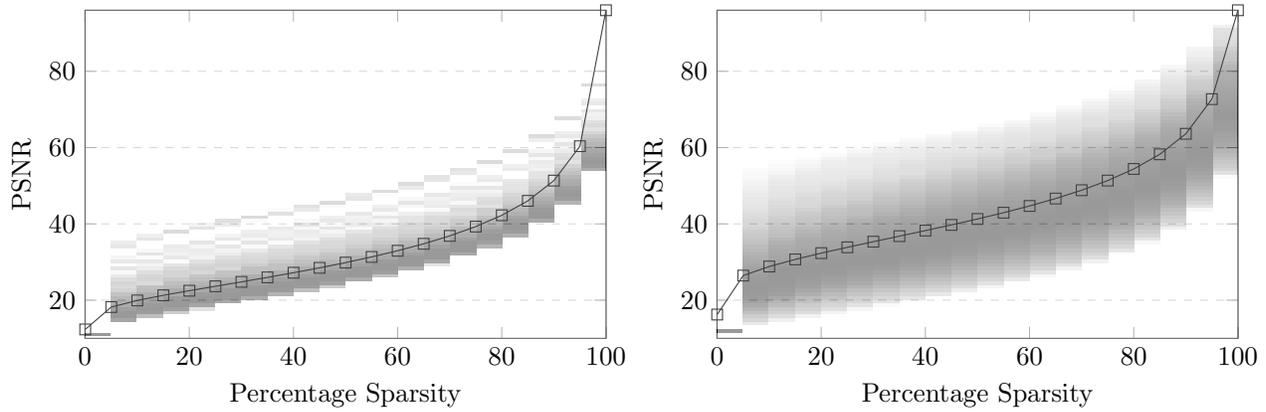

    \centering        
    \begin{subfigure}{0.49\columnwidth}
        \includegraphics[width=\columnwidth]{figures/graph_sparsity_plot.tex}
        \caption{Dead leaves data set}
    \end{subfigure}%
    \begin{subfigure}{0.49\columnwidth}
        \includegraphics[width=\columnwidth]{figures/graph_sparsity_plotNat.tex}
        \caption{Natural scene data set}
    \end{subfigure}
    \caption{Sparse image reconstruction in the Discrete Cosine Transforms with the average PSNR plotted and the PSNR histograms overlaid on the graph as a grey scale image for each different sparsity percentage.}
    \label{fig:sparse_summary} 
\end{figure}

The PSNR was calculated between the generated sparse image and the original image in the data set. Example images at different sparsity is shown in Figure~\ref{fig:sparse_natural} and Figure~\ref{fig:sparse_deadleaves}. The sparsity performance graphs for each data set is shown in Figure~\ref{fig:sparse_summary} with the histogram distribution for each sparsity level overlaid. In the figure it is clear that the Dead leaves data set persistently ranges at the bottom of performance of the natural scene data set and is, therefore, less sparse on average.

\begin{figure}[h!t]
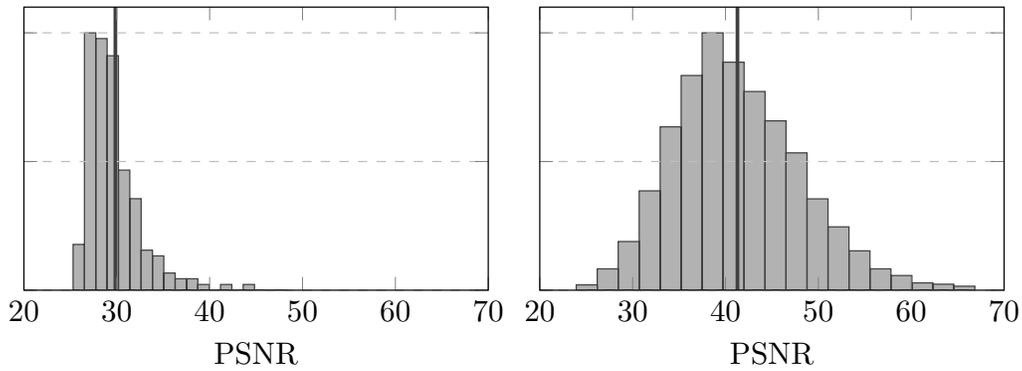
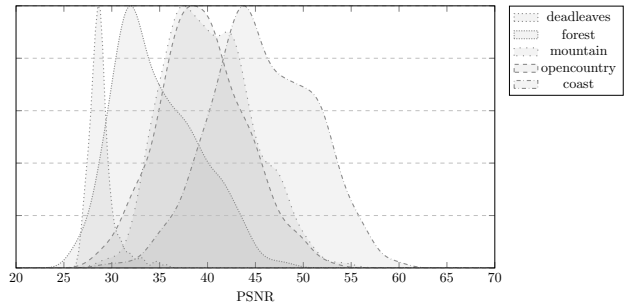

    \centering        
    \begin{subfigure}{0.4\columnwidth}
        \includegraphics[width=\columnwidth]{figures/graph_sparsity_hist.tex}
        \caption{Dead leaves data set}
    \end{subfigure}%
    \begin{subfigure}{0.4\columnwidth}
        \includegraphics[width=\columnwidth]{figures/graph_sparsity_Nathist.tex}
        \caption{Natural scene data set}
    \end{subfigure}
    \begin{subfigure}{0.8\columnwidth}
    \centering
        \includegraphics[width=0.6\columnwidth]{figures/graph_sparsity_cat.tex}
        \caption{Normalised Distributions of different categories indicating difference in image sparsity.}
        \label{fig:sparsity_cat}
    \end{subfigure}
    \caption{The histogram of PSNR values for all images in each respective data set at a 50\% sparsity reconstruction using the Discrete Cosine Transforms}   
    \label{fig:sparse_summary_histogram} 
\end{figure}

In Figure~\ref{fig:sparse_summary_histogram} a single PSNR histogram is shown for 50\% sparsity from which it is clear that the Dead leaves data set covers the bottom part of the Natural Scene database. A worse PSNR value for the same amount of sparse coefficients  indicate that the specific image is less sparse than another. Using this concept the various categories in the natural scene data set and the dead leaves data set can be plotted in a Figure~\ref{fig:sparsity_cat}. The distribution created indicate the various sparsity categories. It is observed that the forest category achieve almost equivalent performance to the dead leaves category where the landscape categories outperform the average PSNR of the dead leaves significantly. This can be attributed to the sparseness found in the various categories and can be evaluated intuitively by comparing the detail in a forest-like image and the detail in an ocean view. It can be concluded that the dead leaves data set will provide a worst case performance and should be adequate to compare various algorithms without using full natural scene data sets. 

\subsubsection{Sparsity in a focal-plane array}
\label{sec:fpa_sparsity}
In imaging and optics it is common practise to use the spatial frequency response (SFR) as an indication to how much detail can be observed in an image. The degradation of an image based on the SFR is commonly modelled through a Gaussian filtering process. Applying Gaussian filtering to an image reduces the amount of high frequencies observable in the image. It is tempting to say, "high detailed images are less sparse than low detailed images". This statement can easily be disproved by creating an image with a single 2-dimensional high-frequency sine-wave. In the Fourier domain such a signal will only require two coefficients. Applying a Gaussian filter to the same image will increase the required Fourier coefficients to represent the new image. Optical images taken from a natural scene is not sparse and thus by adding the assumption that the image is of a natural scene, one can comment that a high detailed image is less sparse. 

\begin{figure}[h!t]
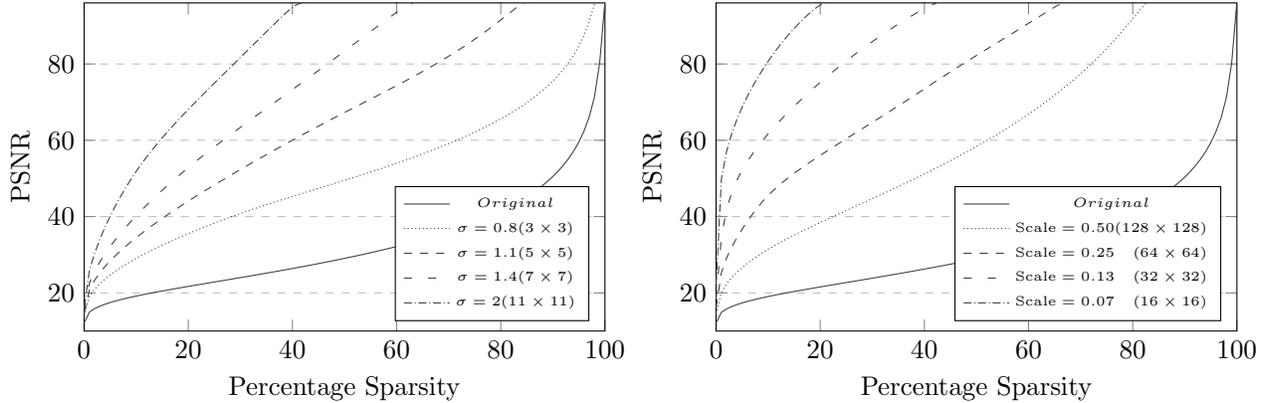

    \centering        
    \begin{subfigure}{0.49\columnwidth}
        \includegraphics[width=\columnwidth]{figures/graph_sparsity_gaus.tex}
        \caption{Gaussian Filtering versus image sparsity}
        \label{fig:sparsity_gaus}
    \end{subfigure}%
    \begin{subfigure}{0.49\columnwidth}
        \includegraphics[width=\columnwidth]{figures/graph_sparsity_down.tex}
        \caption{focal-plane array versus Image sparsity}
        \label{fig:sparsity_down}
    \end{subfigure}%
    \caption{Two graphs indicating how reducing the expected SFR influence the image sparsity. A lower expected SFR produces a sparser natural image in the Discrete Cosine Transform basis.}   
    \label{fig:sparsity_reduction_g} 
\end{figure}

A focal-plane array can be simulated by using a large image, applying an appropriate Gaussian filter followed by area averaging (downsampling) to measure the intensity of each pixel. From the previous section we showed how the PSNR changed over the sparsity of an image here the same sparsity effects can be observed through Gaussian filtering and down sampling of images and is shown in Figure~\ref{fig:sparsity_reduction_g}. It is observed that higher filtering and down sampling require less coefficients to reconstruct the complete image. Using the sparsity measure and PSNR values we can estimate the performance of a rosette imager by the size of a focal-plane array. 

\subsection{Experimental Protocol}
\label{sec:protocols}
We conducted two experiments in order to assess the performance of the Rosetta Imager. The experiments use group comparison through Bayesian Estimation as presented by Kruschke\cite{Kruschke2013}. From Section~\ref{sec:sparsity} the PSNR distributions validate the usage of a student t-distribution for the model approximation and with no previous results, the priors are taken as uniform (See Appendix~\ref{app:student}). 

Rather than using p-values or confidence interval for evaluating the null hypothesis of effect size, we instead use Bayesian estimation with a Highest Density Interval of $95\%$ measured on the effect size. Thus the relative performance of two algorithms are evaluated by approximating the posterior distribution of the effect size via Markov Chain Monte Carlo (MCMC). Typically negative effect size values support the first algorithm and a positive effect size values suppor the second algorithm. However from a Bayesian perspective the algorithm is assumed to perform better if the 95\% Highest Density Interval (HDI) of the effect size are strictly positive. 

The first experiment evaluate the optimal rosette pattern for limiting rotational frequency of 2500~Hz against three other possible patterns. The first pattern, (11, 4) is a previously published pattern, although the purpose of this pattern was only to demonstrate image formation using a rosette scanner and compressed sensing. The second pattern is selected by limiting the maximum rotational frequency to 1000~Hz. The third pattern, (56, 99), was randomly selected from the top 10 patterns found through the optimisation process. Each pattern is used on both data sets outputting a set of PSNR values with related averages and standard deviations. These outputs are then used to perform Bayesian estimation.

The second experiment compare the performance of a focal-plane array~(FPA) to the optimal rosette pattern. The focal-plane array is simulated by down sampling the image to the appropriate size, followed with a linear up scaling to allow for equivalent PSNR measurements. The experiment is only performed using the dead leaves data set. Comparing a single sized FPA to the optimal rosette pattern would not provide any insight and therefore a range of FPA sizes are tested. Using the effect size and the 95\% HDI the expected probability that an image from the FPA will be better than the rosette imager can be estimated.

\section{Results}
\label{sec:results}
\subsection{Scanning Pattern Results}
\begin{figure}[h!t]
    \centering
    \begin{subfigure}{.24\columnwidth}
        \centering
        \includegraphics[width=0.9\linewidth]{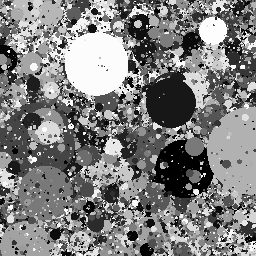}
        \caption*{\phantom{.}}
    \end{subfigure}%
    \begin{subfigure}{.24\columnwidth}
        \centering
        \includegraphics[width=0.9\linewidth]{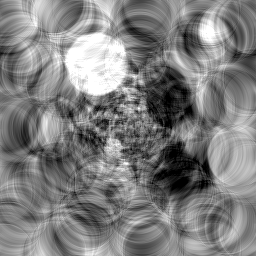}
        \caption*{PSNR: 12.9829}
    \end{subfigure}%
    \begin{subfigure}{.24\columnwidth}
        \centering
        \includegraphics[width=0.9\linewidth]{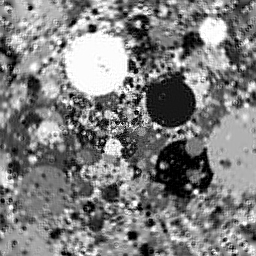}
        \caption*{PSNR: 16.0587}
    \end{subfigure}%
    \begin{subfigure}{.24\columnwidth}
        \centering
        \includegraphics[width=0.9\linewidth]{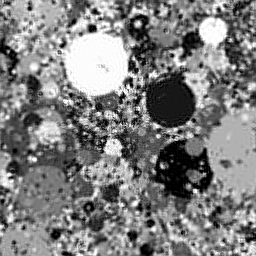}
        \caption*{PSNR: 16.3123}
    \end{subfigure}
    \begin{subfigure}{.24\columnwidth}
        \centering
        \includegraphics[width=0.9\linewidth]{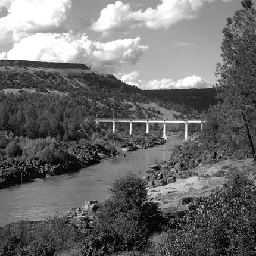}
        \caption{. \\ . \\ Original}
        \label{fig:result_original}
    \end{subfigure}%
    \begin{subfigure}{.24\columnwidth}
        \centering
        \includegraphics[width=0.9\linewidth]{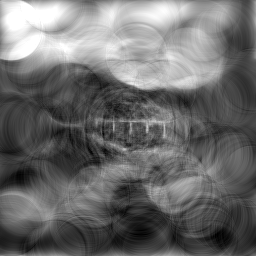}
        \caption{\phantom{.} \\PSNR: 18.7237 \\ (11 4)}
        \label{fig:result_11_4}
    \end{subfigure}%
    \begin{subfigure}{.24\columnwidth}
        \centering
        \includegraphics[width=0.9\linewidth]{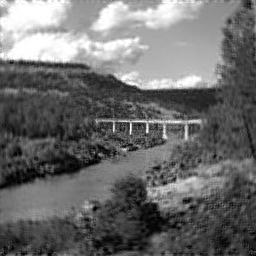}
        \caption{\phantom{.} \\PSNR: 23.2185 \\ (76 37)}
        \label{fig:result_76_37}
    \end{subfigure}%
    \begin{subfigure}{.24\columnwidth}
        \centering
        \includegraphics[width=0.9\linewidth]{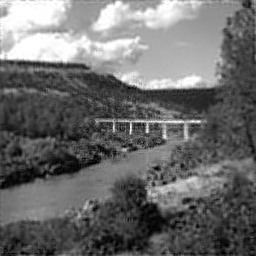}
        \caption{\phantom{.} \\PSNR: 23.3288 \\ (8 99)}
        \label{fig:result_8_99}
    \end{subfigure}%
    \caption{Example images generated by using the various scanning patterns.}
    \label{fig:result_examples}
\end{figure}

The results of the four patterns are shown in Table~\ref{tab:results} with example images in Figure~\ref{fig:result_examples}.
\begin{table}[ht]
    \centering
    \caption{rosette imager pattern evaluation results}
    \label{tab:results}
\begin{tabular}{|c|c|c|c|c|c|}
Pattern & Frequency 1 & Frequency 2 & Probe Size & Dead leaves & Natural Scene \\
(n, m) & (Hz) &(Hz) & (pixels) &(Average PSNR) & (Average PSNR) \\
\hline
(11 4)  &  100 & -175 & 31   & 13.63(1.05) & 20.22(2.84)\\
(76 37) &  925 & -975 & 2.25 & 17.36(1.19) & 25.58(3.76)\\
(56 99) & 2475 & 1075 & 2.25 & 17.67(1.25) & 25.90(3.86)\\
( 8 99) & 2425 & 2275 & 2.25 & 17.67(1.24) & 26.00(3.90)\\
\end{tabular}
\end{table}

First (11 4) and (8 99) is compared. A comparison between averages in PSNR show that (8, 99) outperform (11 4) on both data sets and the difference far exceeds the standard deviations. An example comparison between the two patterns can be made visually in Figure~\ref{fig:result_11_4} and Figure~\ref{fig:result_8_99}.

\begin{figure}[ht]
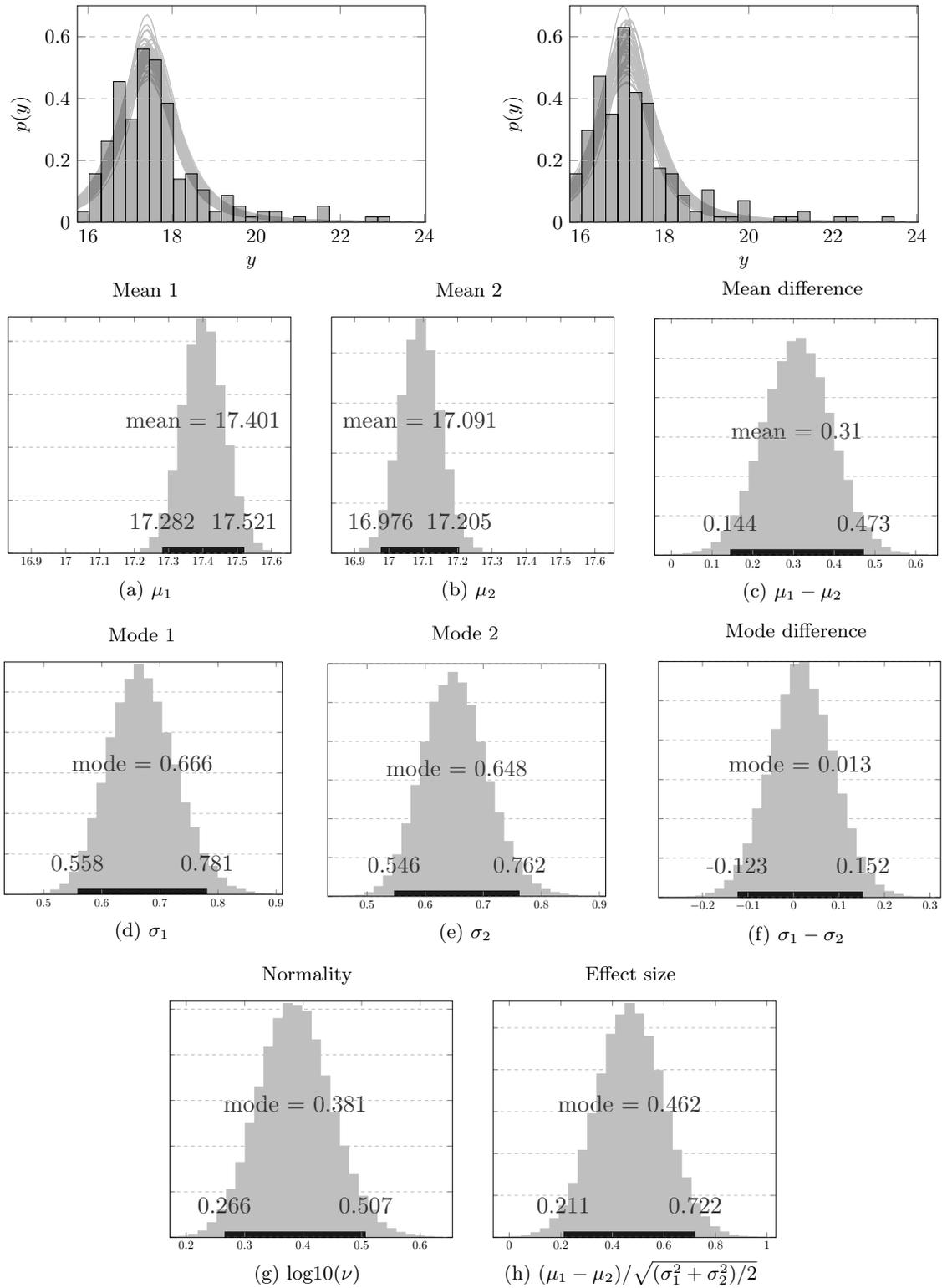

     \centering
    \begin{subfigure}{0.45\columnwidth}
    \centering
        \caption*{Data Group 1 (76 37) w. Post. Pred. ($\mathrm{N}_{Group 1}=200$)}
        \includegraphics[width=0.9\columnwidth]{./figures/BI876_posterior1.tex}
    \end{subfigure}
    \begin{subfigure}{0.45\columnwidth}
    \centering
        \caption*{Data Group 2 (8 99) w. Post. Pred. ($\mathrm{N}_{Group 2}=200$)}
        \includegraphics[width=0.9\columnwidth]{./figures/BI876_posterior2.tex}
    \end{subfigure}
    
    \begin{subfigure}{0.3\columnwidth}
        \centering
        \caption*{Mean 1}
        \includegraphics[width=0.9\columnwidth]{./figures/BI876_mean1.tex}
        \caption{$\mu_1$}
    \end{subfigure}%
    \begin{subfigure}{0.3\columnwidth}
        \centering
        \caption*{Mean 2}
        \includegraphics[width=0.9\columnwidth]{./figures/BI876_mean2.tex}
        \caption{$\mu_2$}
    \end{subfigure}%
    \begin{subfigure}{0.3\columnwidth}
        \centering
        \caption*{Mean difference}
        \includegraphics[width=0.9\columnwidth]{./figures/BI876_mean_diff.tex}
        \caption{$\mu_1-\mu_2$}
    \end{subfigure}%
    
    \begin{subfigure}{0.3\columnwidth}
        \centering
        \caption*{Mode 1}
        \includegraphics[width=0.9\columnwidth]{./figures/BI876_mode1.tex}
        \caption{$\sigma_1$}
    \end{subfigure}%
    \begin{subfigure}{0.3\columnwidth}
        \centering
        \caption*{Mode 2}
        \includegraphics[width=0.9\columnwidth]{./figures/BI876_mode2.tex}
        \caption{$\sigma_2$}
    \end{subfigure}
    \begin{subfigure}{0.3\columnwidth}
        \centering
        \caption*{Mode difference}
        \includegraphics[width=0.9\columnwidth]{./figures/BI876_std_diff.tex}
        \caption{$\sigma_1 - \sigma_2$}
    \end{subfigure}
    
    \begin{subfigure}{0.3\columnwidth}
    \centering
    \caption*{Normality}
        \includegraphics[width=0.9\columnwidth]{./figures/BI876_norm.tex}
        \caption{$\mathrm{log10}(\nu)$}
    \end{subfigure}%
    \begin{subfigure}{0.3\columnwidth}
    \centering
    \caption*{Effect size}    
        \includegraphics[width=0.9\columnwidth]{./figures/BI876_effect_size.tex}
        \caption{   $ (\mu_1 - \mu_2)/\sqrt{(\sigma_1^2 + \sigma_2^2)/2}$}
    \end{subfigure}    
    \caption{The results of the Bayesian estimation between (76 37) and (8 99). The 95\% HDI of the effect size indicate that (8 99) is statistically better than (76 37).}
    \label{fig:bayes_8_99_76_37}
\end{figure}

The second evaluation is between (8 99) pattern and (76 37) pattern. Comparing the means of the PSNR of each data set, (8 99) is better than (76 37) although the standard deviations are large compared to the difference in means. A visual inspection can be performed between Figure~\ref{fig:result_76_37} and Figure~\ref{fig:result_8_99} and only slight degradation's can be observed between the dead leaves patterns. This difference might only be specific to the image example. To dig deeper in the comparison the Bayesian estimation is shown in Figure~\ref{fig:bayes_8_99_76_37}. 

\FloatBarrier

In the first row of the figure the estimated posterior probability distribution of the PSNR values are overlaid with the histogram data of the sampled data. The second row show the posterior distributions for the $\mu$ parameter and the difference of means indicate that the difference are statistically significant. It is interesting to note how these means differ from the averages shown in Table~\ref{tab:results}. If the data is roughly symmetric, the comparison of the $\mu$ parameter can be adequate. The third row provide the mode or somewhat the data variance. The difference in modes according to the 95\% HDI is not significant. The lower left graph provide the distribution of the normality parameter, where small values indicate large tailed distributions. The posterior distribution is relatively narrow indicating the consistent accommodation of outliers within the data. The last posterior distribution show the credible effect size of the two data groups. The 95\% HDI of the effect size supports the hypothesis that group 2 is significantly higher than group 1. The Bayesian estimation shows that the 95\% HDI of the effect size is not shared between the two patterns and therefore it is expected that (8 99) performs better than the (76 37) pattern.

\begin{figure}[h!t]
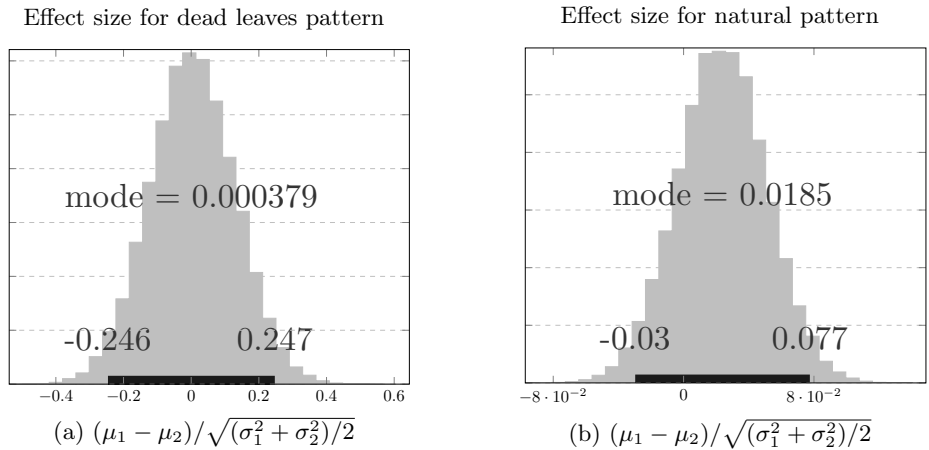

    \centering
    \begin{subfigure}{0.4\columnwidth}
        \centering
        \caption*{Effect size for dead leaves pattern}
        \includegraphics[width=0.8\columnwidth]{./figures/BI856_effect_size_dead.tex}
        \caption{$(\mu_1 - \mu_2)/\sqrt{(\sigma_1^2 + \sigma_2^2)/2}$}
    \end{subfigure}%
    \begin{subfigure}{0.4\columnwidth}
        \centering
        \caption*{Effect size for natural pattern}
        \includegraphics[width=0.8\columnwidth]{./figures/BI856_effect_size_natural.tex}
        \caption{$(\mu_1 - \mu_2)/\sqrt{(\sigma_1^2 + \sigma_2^2)/2}$}
    \end{subfigure}    
    \caption{The effect sizes of Bayesian estimation between (56 99) and (8 99) on the dead leaves data set and natural scene data set. The 95\% HDI indicate that there is no significant difference between the two patterns.}
    \label{fig:bayes_8_99_56_99}
\end{figure}

The third evaluation, between (8 99) pattern and (56 99) pattern, is somewhat more tricky. Evaluating the PSNR averages (8 99) performs equivalent to that of (56 99) on the dead leaves pattern but much better on the natural data set. A Bayesian estimation is performed on the dead leaves data set and the natural scene data set with only the output of the effect sizes shown in Figure~\ref{fig:bayes_8_99_56_99}. Taking the 95\% HDI, the difference in performance is not statistical significant and we can therefore not say that the one pattern is better than the other.
\FloatBarrier

\subsection{Focal-plane Array Comparison Results}
\begin{figure}[h!t]
    \centering
    \begin{subfigure}{0.5\columnwidth}
        \includegraphics[width=\columnwidth]{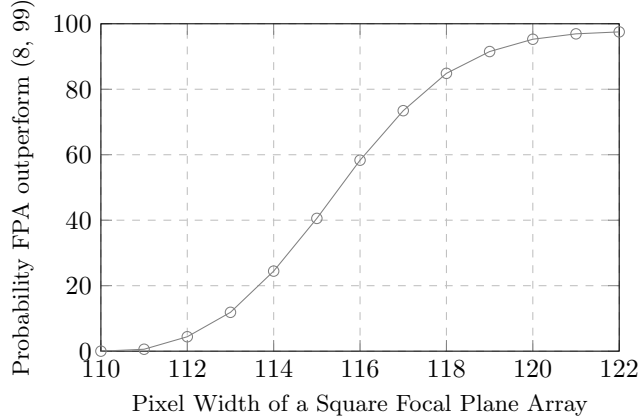}
    \end{subfigure}
    \caption{The probability according to the 95\% HDI that the specific sized FPA will outperform the rosette imager on a sampled image.}
    \label{fig:bayes_fpa_size}    
\end{figure}

The results for comparing the range of focal-plane arrays to the (8 99) pattern is shown in Figure~\ref{fig:bayes_fpa_size}. It can be seen from the graph that an FPA of size $114\times114$ will only produce an image $30\%$ of the time with a higher PSNR value than the rosette imager. From the analysis the rosette imager will produce an image 95\% of the time equivalent to a $111\times111$ up to $121\times121$ FPA. Uniformly sampling the imaging area $12974$ samples can be used in image reconstruction and thus an FPA with equivalent number of samples rounded up would be a $114\times114$ FPA.

\FloatBarrier

\section{Discussion}

The designed rosette imager is based on a rosette scanner and the Fourier Domain Regularisation Inversion algorithm. The imager use a single sensor with an instantaneous field of view equal to 2.25~pixels; It samples at 510~kHz to output a $256\times256$ image at 25~frames per second. Furthermore, the wedge rotates at 2475Hz, and the mirror rotates at 2275~Hz. The set-up produced images of equivalent quality as focal-plane arrays of size $111\times111$ to $121\times121$.

We expected that the rosette imager would perform just as well as the focal-plane array. The performance is due to the signal sparsity requirement in compressed sensing theory. In the exploration of image sparsity we have also shown that natural images are not sparse. The development of an imager using a single detector scanning system 
compared to a focal-plane array allow for different trade-offs (such as image quality versus number of samples needed). Further modelling of a rosette imager will better establish these trade-off parameters. We note that the same rosette pattern can be achieved by replacing the rosette scanner with a Risley prism.

The image sparsity investigation highlighted the effect of different image categories on the expected PSNR and image sparsity itself. Furthermore, the investigation indicates that the performance of algorithms are dependant on the set of testing images. Preferably algorithms should be compared using the same set of images. It is also suggested that using only the dead leaves pattern for comparison should be adequate for a large range of algorithms in image formation. The PSNR measure and image sparsity was shown to provide estimates of the spatial frequency response of the imager. These measurements can likely assist evaluations in the laboratory. Finally, we note that advanced statistical techniques such as Bayesian estimation will benefit the evaluation of image formation and image processing algorithms.

\appendix    

\section{Equations}
\subsection{Rosette scanner governing equations}
\label{app:rosette}
The wedge typed device can be described by the following design equations\cite{Tajime1980}:
\begin{equation*}
\phi_2 \approx \frac{\phi_1(n-1)d_1}{2d_2}
\end{equation*}
with $\phi_1$ the vertical angle of the wedge, $\phi_2$ the canted angle of the secondary mirror, $n$ is the refractive index of the wedge, $d_1$ is the distance of wedge from the sensor and $d_2$ is the distance of the mirror from the sensor. The locus of the IFOV is given by: 
\begin{equation*}
    x = \phi_2 \left( \cos(2\pi f_1 t) -\cos(2\pi f_2 t) \right)
\end{equation*}
\begin{equation*}
    y = \phi_2 \left( \sin(2\pi f_1 t) -\sin(2\pi f_2 t) \right)
\end{equation*}
Furthermore, it is required that the rosette scanning pattern is repeated at a specified time interval $T$; which also relate to the specified frame rate of the imager $1/T$. The frame rate in conjunction with rotating elements are constrained by:
\begin{equation*}
    T = \frac{m}{f_1} = \frac{m-n}{f_2} = \frac{n}{f_1 - f_2}
\end{equation*}
with $n,m \in \mathbb{Z}$.

\subsection{Peak-Signal to Noise Ratio Equations}
\label{app:psnr}
Given an image $I$ of height $H$ and width $W$ such that $I(m, n) \mapsto i_{m,n} \in {0, \ldots,2^{16} - 1}$ with $|m| \leq H-1$, $|n| \leq W-1$ and $m,n \in \mathbb{Z}$. The Mean Square Error between a reference image $I_{R}$ and a test image $I_{T}$ is given be
\begin{equation*}
MSE(I_{T}, I_{R}) = \frac{1}{HW} 
\sum_{m=0}^{H-1}\sum_{n=0}^{W-1} \left[ I_{T}(m,n) - I_{R}(m,n)\right]^2
\end{equation*}
The Peak Signal-to-Noise Ratio~(PSNR) in decibel~(dB) is defined as
\begin{equation}
    \label{eqn:PSNR}
    PSNR(I_{T}, I_{R}) = 
    \begin{cases}
    20\cdot \log_{10}\left( \frac{2^{B} - 1}{\sqrt{MSE(I_{T}, I_{R})}}\right) &  MSE(I_{T}, I_{R}) \ge 1\\
    20\cdot \log_{10}\left( 2^{B} - 1 \right) & Otherwise
    \end{cases}
\end{equation}

\subsection{Student t-distribution}
\label{app:student}
The student t-distribution is given by
\begin{equation*}
    \text{StudentT}(y|\nu,\mu,\sigma) = \frac{\Gamma\left((\nu + 1)/2\right)}      {\Gamma(\nu/2)} \ \frac{1}{\sqrt{\nu \pi} \ \sigma} \ \left( 1 + \frac{1}{\nu} \left(\frac{y - \mu}{\sigma}\right)^2 \right)^{-(\nu + 1)/2} \!
\end{equation*}
with $\nu$ as the normality parameter, $\mu$ as the mean, $\sigma$ as standard deviation and $y$ as the data point.

The effect size is calculated by:
\begin{equation*}
    \text{Effect Size} = \frac{\left( \mu_1 - \mu_2 \right) }{\sqrt{\frac{1}{2}\left( \sigma_1^2 + \sigma_2^2\right)}}
\end{equation*}

 

\bibliography{references}
\bibliographystyle{spiebib}

\end{document}